\documentclass[3p,twocolumn]{elsarticle}

\usepackage{amsmath}
\usepackage{xspace}
\usepackage{epsfig}
\usepackage{graphicx}
\usepackage{epstopdf}
\usepackage{amssymb}
\usepackage{color}
\usepackage{subfigure}
\graphicspath{{./}}
\newtheorem{thm}{Theorem}
\newtheorem{theorem}{Theorem}
\newtheorem{lemma}[thm]{Lemma}
\newproof{proof}{Proof}
\newtheorem{example}[thm]{Example}
\newtheorem{definition}[thm]{Definition}
\newtheorem{corollary}[thm]{Corollary}

\newcommand {\lieG} {\mathbf{G}\xspace}

\newcommand {\lieg} {\mathfrak{g}\xspace}

\newcommand {\controlValueSet} {\mathbf{V}}
\newcommand {\controlSet} {\mathcal{V}}
\newcommand {\targetSet} {\mathcal{T}}

\newcommand {\deltatx}{\Delta t^{h}(x)\xspace}

\begin{document}

\begin{frontmatter}



\title{ Dynamic Programming and Viscosity  solutions for
 the optimal control of quantum spin systems}

\author[anu]{Srinivas Sridharan \corref{cor1}}
\ead{srinivas.sridharan@anu.edu.au}
\address[anu]{Department of   Engineering, Australian
    National University, Canberra, ACT 0200,   Australia.} 
\cortext[cor1]{Corresponding author.  Tel: +61  2 612 58827} 

\author[anu]{Matthew R.~James}

\begin{abstract}
The purpose of this paper is to describe the application of the notion of viscosity  solutions to solve the Hamilton-Jacobi-Bellman   (HJB) equation  associated with an important class of optimal control problems for quantum spin systems. The HJB equation that arises in the systems of interest  is a first order nonlinear partial differential equation   defined on a Lie group. 
 We employ  recent extensions of the theory of  viscosity solutions  from Euclidean space to Riemannian manifolds to interpret possibly non-differentiable solutions to this equation.
  Results from differential topology on the triangulation of manifolds are then used to develop a  finite difference approximation method, which is shown to converge using viscosity solution methods. An example is provided to illustrate the method.
 \end{abstract}


\end{frontmatter}

%

\section{Introduction}

Recently, there has been considerable attention directed at the problem
of obtaining time optimal trajectories for open loop control of
quantum spin systems \cite{N.2001,khaneja-2003,Dirr2006,carlini2006toq}.
These problems arise from applications which include NMR
 spectroscopy (to produce a time optimal trajectory),
 and the optimal construction of quantum
circuits \cite{nielsen:062323,nielsen2006qcg} (to minimize the
number of logic gates required to construct a desired unitary
transformation). These spin systems have the mathematical structure of a bilinear right
invariant system on the special unitary group. Owing to the importance of the applications, there have been  various approaches to solving these problems which utilize  Lie theoretic arguments \cite{khaneja-2003,dalessandro2001oct}, calculus of variations \cite{carlini2006toq,boscain:062101}  and Dynamic Programming \cite{sjcdc2008,smjpra2008}.

In the dynamic programming approach, under appropriate regularity assumptions,  the optimal cost function (value function)  is the solution to  a Hamilton-Jacobi-Bellmann (HJB) equation  \cite{kirk2004oct,bertsekas1995dpa,bryson1975aoc}. For many problems of interest this  value function can be demonstrated to be  non-differentiable. Hence there is the need for a more general notion of a solution to such PDEs. A popular and successful concept of such a weak solution of nonlinear PDEs  is the well studied  theory of viscosity solutions  \cite{Bardi, Crandall1984} on Euclidean spaces. Because the quantum spin problem leads to a HJB equation defined on a Lie group,  we use extensions of the viscosity solution theory  to Riemannian manifolds  \cite{azagra2005naa,Azagra2006,mantegazza2002hje} in order to   interpret  the  solutions of this  equation.
For a detailed introduction to this topic we refer the reader to \cite{Bardi,lions1982generalized} and references contained therein.

In this article we build up the components required for a rigorous application of viscosity solution theory on manifolds for quantum systems. This commences with an explanation of a discretization method  based on the triangulation of manifolds  \cite{munkres1973edt} to solve the HJB equation for the optimal spin control problem. We then use viscosity solution concepts to prove the convergence of the solution obtained by this triangulation-based discretization scheme to the solution of the original HJB equation.

The structure of this article is as follows. We begin by describing the quantum spin control problem in Sec \ref{sec:siam:sysDesc}. This is followed in Sec \ref{sec:siam:regularityValfun} by a study of the regularity properties of the value function which play an important role in the solution of the associated HJB equation. After motivating the need for more generalized solutions of the HJB equation using an example system with a  non-differentiable value function, we explain
the use of the notion of viscosity solutions on Lie groups  in Sec \ref{sec:siam:viscsoln}.  Results pertinent to the existence and uniqueness of such solutions are recalled from relevant literature and modified to the framework of the problems introduced.  In order to solve these optimal control problems numerically we make use of the notion of triangulation of the  group on which the system evolves. This concept and the proofs of convergence of the approximations to the actual solution using viscosity solution notions are introduced and developed in Sec \ref{sec:siam:discretization}. In Sec \ref{sec:siam:simulation} the ideas developed are then applied to solve an example control problem on $SU(2)$ for which sample optimal trajectories and the value functions from the simulations are obtained. We conclude with comments and  possible extensions in Sec \ref{sec:siam:conclusion}.

\section{Problem Description}\label{sec:siam:sysDesc}
\subsection{System Description}\label{subsec:siam:sysDesc}
In this section we introduce a mathematical model arising in applications of the open loop control of quantum systems.
Given a compact connected matrix Lie group $\lieG=SU(2^n)$ with an associated Lie algebra $\lieg=\mathfrak{su}(2^n)$ and
smooth, right invariant vector fields $X_1,....X_m$ in $\lieg$, let the
evolution of the system be given by
\begin{align}
\frac{dU}{dt} &= \{\sum_{k=1}^m v_k(t) X_k\} \,U,\qquad U(0)=U_0,  \label{eq:siam:System}\\
\qquad U, U_0\,\in\,\lieG. \nonumber
\end{align}
Here $X_1\,,\,\ldots\,,\,X_m$ are termed control vector fields. The $v_i$ are elements of the control signal $v$ which belongs to  the class of piecewise continuous functions $\controlSet$ with their range belonging to a compact subset $\mathbf{V}$ of the real $m$-dimensional Euclidian space $\mathbb{R}^m$, containing the origin. Without loss of generality we may consider $\mathbf{V}$ to be the unit hypercube in $\mathbb{R}^m$ around the origin.  We assume that  the Lie algebra
generated by the set $\{X_1,\,...,\,X_m \}$, using repeated Lie bracketing operations of all orders, is $\lieg$. 
We denote the right hand side of the Equation~\eqref{eq:siam:System} above by $f(U,v)$.  Given a control signal  $v$ and an initial point $U_0$, the
solution to Eq~\eqref{eq:siam:System} at time $t$ is denoted by
$U(t;v,U_0)$. We denote by $\targetSet$ a compact set in $\lieG$ with smooth closed boundary $\partial \targetSet$. This set is the target set that we wish the system to reach.

Now, the following properties are satisfied at every point $x \in \lieG$:
\begin{itemize}
\item the system dynamics $f(\cdot)$ is driftless,
\item if $\mathcal{X}:=f(x,u)$ can be generated by a certain control $u\in\controlValueSet$, then  $-\mathcal{X}$ can also be generated using another element of the control set (in this case it would be $-u$),
\item the dimension of the vector space  at $x$, generated by the set of vector fields after all possible bracketing operations, has the dimension equal to that of the group.
\end{itemize}
Hence we have from \cite[Prop. 3.15]{nijmeijer1990ndc} that the time to get from any point on the group, to the identity element is bounded.  Thus the entire group is reachable from the identity and hence the problems dealt with in the next section are well defined.

\subsection{Problem Formulation}\label{subsec:siam:probForm}
A large class of problems in the control of quantum spin systems and quantum circuit synthesis can be recast in terms of an optimal control problem with the following value function (optimal cost function):
\begin{align}
  S(U_0) &= \mathop {\inf
  }\limits_{v\in\controlSet} \left \{ {\int\limits_0^{t_{U_0}(v) } {\ell(U(s;v,U_0),v(s))\,e^{-\lambda\,s}\,} ds}\right \},\label{eq:siam:ValueFunctionGeneral}
  \end{align}
  where   $\ell\,:\,\lieG\,\times\,\mathbf{V} \rightarrow\,\mathbb{R}$ is continuous and $\lambda > 0$ is a real valued discount factor. Here $t_{U_0}(v)$ denotes the time to reach the set $\partial \targetSet$ starting from $U_0$ using control $v$. Note that under the assumption that the cost $\ell(.)$ has a lower bound which is positive,  we can then assume without loss of generality that the minimum of $\ell$ over $\lieG\, \times\, \mathbf{V}$ is~$1$.
We now proceed to study some properties of the value function defined above.

\section{Regularity of the Value Function}\label{sec:siam:regularityValfun}
We begin by introducing certain quantities which will be used in the study of the regularity properties of the value function.  Define the minimum time function
\begin{align}
T(U):=\inf \limits_{v\in\controlSet}\{t_{U}(v)\},\qquad U\in \lieG, \label{eg:siam:defMinTimFun}
\end{align}
which is the infimum of the time taken to reach the target set from a starting point $U$.
The set of points, termed the reachable set, from which the desired target set may be reached in time
$\tau$ is defined as
$$
R(\tau):=\{U\in \lieG \,|\;T(U)<\tau\}.
$$
A system is said to be \textit{small time controllable} on a set
$\targetSet$ (denoted by STC$\targetSet$)  if
\begin{align}
\targetSet\subseteq \mathop{int}(R(t)),\quad \forall\; t>0.
\label{eq:siamWriteup:STCT}
\end{align}
Note that the assumption of small time controllability to the target set is required to obtain some of the results in this section and holds for the examples studied in this paper (and may be verified for any system under consideration). 

We now introduce some results on the  regularity of the value function whose proofs proceed along the lines of the arguments used in \cite{Bardi} with suitable modifications due to the Lie group setting. 
\begin{lemma} [ Proposition 1.2 ($\S$ 4) \\ \cite{Bardi}]\label{lem:siamWriteup:ContinuityofValueFunctionUsingSTCT}
If the system is small time controllable on the set $\targetSet$
then the value function is continuous in some open set containing
the boundary  $\partial \targetSet$ of the set.
\end{lemma}
\begin{lemma}  \label{lem:siamWriteup:SisBC}Given a system evolving on a connected, compact Lie group, with dynamics ( Eq~\eqref{eq:siam:System}) such that $f$, $\ell$, $S$  satisfy the following conditions:
\begin{enumerate}
  \item $f:\, \lieG \,\times
  \,\controlValueSet\,\rightarrow\,T\lieG$ is continuous.
  \item $\ell\,:\, \lieG\,\times
  \,\controlValueSet\,\rightarrow\,\mathbb{R}$ is continuous.
  \item $S$ is continuous on some open set containing $\partial\targetSet$
\end{enumerate}
Then $S$ is bounded and continuous on $\lieG$.
\end{lemma}
\begin{proof}
This result consists of  two cases of the value function.
\begin{itemize}
\item $\lambda = 0$: As indicated in Sec.\ref{subsec:siam:sysDesc}, the time to get from any point on the group, to the identity element is bounded.  
This along with the bounds on $\ell$,  imply  that  the value function is bounded. 
\item $\lambda>0$: In this case  the boundedness follows directly from the bounds on $\ell$ and the exponential decay factor $\lambda>0$. 
\end{itemize}
In both cases, the continuity proofs proceed along the lines of  \cite[Prop 3.3 ($\S$ 4) ]{Bardi} with suitable modifications for the Lie group setting.
\end{proof}

\begin{example} [Property  of Value Function]  \label{eg:siam:not-dbl}\end{example}
Consider the following system defined on the special unitary group
 $\mathbf{G}=SU(2)$:
\begin{align}
   \dot{U}  &=  [I_z  + v I_x ]\, U,\, \label{equation:siam:SUsystem}  \\
   U(0)  &= U_0, \nonumber
\end{align}
where $v:[0\,,\,\infty)\rightarrow (-\infty\,,\,\infty)$ is a piecewise continuous control signal.
Here $I_x$ and $I_z$ are given by:
\begin{align}\label{eg:siam:mathconvention}
I_x&= \frac{-j}{2}\,\left(
  \begin{array}{cc}
    0 & 1 \\
    1 & 0 \\
  \end{array}
\right),\quad I_z = \frac{-j}{2} \,\left(
  \begin{array}{cc}
    1 & 0 \\
    0 & -1 \\
  \end{array}
\right).
\end{align}
Note that in Eq~\eqref{eg:siam:mathconvention},  we follow the mathematicians' convention  in which $I_x$ and $I_z$ are skew-Hermitian matrices which belong to the Lie algebra $\mathfrak{su}(2)$ of the group $SU$(2).
The cost function to reach the target set  for this problem  (which in this case is the identity element $I$) is given by
\begin{align}
S(U_0)&= \mathop {\inf}\limits_{v\in \controlSet } {\int\limits_0^{t_{U_0}(v)} {\exp\{-\lambda\,s\}\, ds}}.
\end{align}
We demonstrate below  that this normalized minimum time function
$S\,:\,SU(2)\,\rightarrow\,\mathbb{R}$ is not
differentiable at the points $\pm{I}$.

Any point $P$ in $SU(2)$  can be represented as:
$k_1\,\exp(\alpha\,I_z)\,k_2$  where $k_1\,,\,k_2$ are elements of
the Lie subgroup generated by $\exp(I_x)$  and $\alpha
\,\in\,[0\,,\,\pi]$. 
Using \cite{N.2001}, the  cost function
for any point expressed in this form is given by 
$$
\frac{1-\exp\{ -\lambda \alpha\}}{\lambda}.
$$
For this
compact, connected Lie group the exponential mapping (denoted by
$\phi$) is a diffeomorphism from an open set around the origin in
 $\mathfrak{su}(2)$ to an open set around $I$ in
$SU(2)$. Let the $I_z$ axis in
$\mathfrak{su}(2)$ be denoted by $e_1$.
If the function $S$ is differentiable at the
identity element then the function
$\tilde{S}\,:=\,S\,\circ\,\phi\,:\,\mathfrak{su}(2)\,\rightarrow\,\mathbb{R}$
must be differentiable at the origin in $\mathfrak{su}(2)$. Hence
there must must exist a linear function $\eta:\mathbb{R}^3 \rightarrow \mathbb{R}$ s.t
\begin{align}
\mathop {\lim }\limits_{\left\| \varepsilon  \right\| \to 0} \left\{
{\frac{{\tilde{S}(x + \epsilon ) - \tilde{S}(x) - \eta
(\epsilon )}}{{\left\| \epsilon  \right\|}}} \right\} = 0\,,\quad
\mathrm{at}\,\,x\,=\,0. \label{eq:siam:tildesdifferentiable}
\end{align}
Now, consider a line through the origin in $\mathfrak{su}(2)$ along
the $e_1$ axis. Let  $\epsilon$ be either $+ \delta \,e_1$ or
$-\delta \, e_1$ (with $\delta\,>\,0$). The value of $\tilde{S}$ has the following properties at the origin
\begin{align}
\tilde{S}(0) &=0\nonumber\\
\tilde{S}(+\, \delta\, e_1) & =\tilde{S}(-\, \delta \, e_1) = \frac{1-\exp\{-\lambda \delta\} }{\lambda}. \label{eq:siam:splussminus}
\end{align}
In addition, at the origin the function $\eta(\epsilon)$ takes the form $\delta L$ where $$L := e_{1}\cdot D\tilde{S}(0),$$
where $D\tilde{S}$ denotes the differential of the value function.  Hence using this expression for  $\eta(\epsilon)$
and Eq~\eqref{eq:siam:splussminus} in Eq~\eqref{eq:siam:tildesdifferentiable}, it follows that for $\tilde{S}$ to be differentiable at the origin
\begin{align}
\mathop {\lim }\limits_{\delta \to 0} \left\{
{\frac{{\tilde{S}(\pm \delta \epsilon_{1} ) - \delta L}}{{\delta}}} \right\} = 0. \label{eq:siam:diff2}
\end{align}
Thus from Eq~\eqref{eq:siam:splussminus} and the Taylor expansion for $\exp\{\lambda \delta\}$ it follows that the two equations below must 
simultaneously hold in order to ensure differentiability.
\begin{align}
\mathop {\lim }\limits_{\delta \to 0} \left\{
{\frac{ \delta- \delta L+ O(\delta^{2})}{\delta}} \right\} = 0 \label{eq:siam:diff3},\\
\mathop {\lim }\limits_{\delta \to 0} \left\{
{\frac{ \delta+ \delta L+ O(\delta^{2})}{\delta}} \right\} = 0 \label{eq:siam:diff4}.
\end{align}
The first of these requires $L=+1$ and the second requires $L=-1$, leading to a contradiction. Hence $\tilde{S}$ is not differentiable at the origin, 

and therefore the function $S$ is not differentiable at $I$. Similar arguments
hold for the element $-I$ of $SU(2)$. $\Box$

In view of this example, care is needed when interpreting the notion of solutions to the HJB equation since the value functions may not be differentiable. This  motivates our study of the use of 
 non-differentiable weak solutions (more specifically viscosity solutions) to the HJB equation \cite{Bardi}.

\section{Viscosity Solution Theory}\label{sec:siam:viscsoln}
\subsection{Introduction}
The example in the previous  section indicates the need for notions of weak solutions to the HJB  equation associated with the dynamic programming problem. In this section we recall certain recent extensions of   viscosity solution theory to Riemannian manifolds \cite{azagra2005naa,Azagra2006,mantegazza2002hje} and present  them in a form suitable to our study of systems evolving on the special unitary group $\lieG$.   We then  indicate certain results about the existence and uniqueness of such solutions. 

Apart from being useful as  a weak solution to the HJB which we require, the viscosity solution concept will also be used to prove convergence (in a specific sense) of the numerical approximation schemes  to the original problem.  The definitions and results in this article are more generally applicable to Riemannian manifolds, but for our current problem formulation the Lie group setting is sufficiently general. Note that  we use the notation $C^{k}(\Omega)$  for the class of functions whose $k$-th order derivatives are continuous on a set $\Omega$. In what follows $D \phi : T\lieG \to \mathbb{R}$ denotes the differential of a function $\phi:  \lieG \to \mathbb{R}$.

We start by defining the notion of viscosity solutions from  \cite[Chapter 2]{Bardi}
\begin{definition}[Continuous Viscosity Solution]
Given an open domain $\Omega$ in a Lie group $\lieG$, a function $S\,\in\,C(\overline{\Omega})$ is a
viscosity sub(super) solution of the following PDE in $\Omega$
\[
F(U,S,DS) = 0 \label{eq:siamWriteup:PDE}\,,\,U\,\in\,\Omega
 \]
if $\forall\,\phi\,\in\,C^1(\overline{\Omega})$
\[
{F(\bar{U},S,D\phi )} \leq\,(\,\geq\,)\, 0
\]
at every point $\bar{U} \,\in\,\overline{\Omega}$ where $S-\phi$ has a relative
maxima (resp. minima). A function is a viscosity solution iff  it is
both a super and sub viscosity solution.
\end{definition}

\subsection{On the Existence and Uniqueness of Viscosity Solutions}
We proceed to develop certain results regarding the viscosity solution to the HJB equation arising from the associated control problem.  
\begin{lemma}\label{lem:siam:SisaContinViscSoln}
The value function $S$  given by Equation~\eqref{eq:siam:ValueFunctionGeneral} is a continuous viscosity solution of the HJB equation
\begin{align}
F(U,S,DS):=\lambda\, S+H(U,DS)=0,  \label{eq:siam:generalHJB}\\
 U\in \lieG \backslash \targetSet,\quad \lambda >0 \nonumber
\end{align}
 where
$$
H(U,DS):=\mathop {\sup }\limits_{v \in \controlValueSet} \left\{ {
- DS(U) \cdot f(U,v) - \ell(U,v)} \right\}.
$$
 \end{lemma}
\begin{proof}
This follows along the lines of \cite[Prop 3.11 ($\S$ 4)]{Bardi}, with slight modifications.
\end{proof}

Now, in order to prove uniqueness of the viscosity solution, we recall the following lemma.

\begin{lemma}[Corollary 4.10 of \cite{Azagra2006}]\label{lem:siam:generalUniquenessViscResult}
Let $\lieG$ be a compact Lie group and $\Omega$ be a bounded open subset of $\lieG$. Suppose that $F : T^{\ast}\lieG \times \mathbb{R} \rightarrow \mathbb{R}$ is uniformly continuous. Given that $\forall r \geq s$
\begin{align}
\exists \gamma > 0 \,\text{s.t}\quad 
\gamma (r - s) \leq F (x, r, \zeta) - F (x, s, \zeta), \end{align}
where $ x \in \lieG$. There is at most one viscosity solution of the HJB equation $$F(x,S,DS(x))=0.$$
\end{lemma}
We arrive now at the main uniqueness results for the discounted value function.

\begin{corollary}\label{siam:thm:SuniqueVS}
Assume that the value function $S$ (with $\lambda>0$) is continuous. It is the unique viscosity solution to the HJB  
 equation~\eqref{eq:siam:generalHJB} on $\overline{\lieG \setminus \targetSet}$.
\end{corollary}
\begin{proof}
From Lemma \ref{lem:siam:SisaContinViscSoln}, $S$ is a viscosity solution to the HJB equation \eqref{eq:siam:generalHJB}. In addition, this HJB equation  can be  shown to satisfy the conditions in Lemma \ref{lem:siam:generalUniquenessViscResult}. Hence the statement of this corollary follows.
\end{proof}

We now deal with the case of the undiscounted cost, starting specifically with the minimum time function, since it is of special interest in applications. 
\begin{theorem}\label{siam:thm:SuniqueVSmintime}
Assume that the minimum time function $S$  (with $\ell =1$ and $\lambda =0$) is continuous. It is the unique viscosity solution to the HJB 
 equation~\eqref{eq:siam:generalHJB} on $\overline{\lieG \setminus \targetSet}$.
\end{theorem}
\begin{proof}
In this case we have that $S$ is a viscosity solution of the HJB equation
\begin{align}
\mathop {\sup }\limits_{v \in \controlValueSet} \left\{ { -
DS(x)\cdot f(x,v) -1} \right\}
=0, \,\,x \in \lieG.\label{eq:siam:uniquenessOfContinuousValueFunctionsmintime:HJB}
\end{align}
Applying a particular diffeomorphism called the Kruskov transform to $S$
we obtain a new value function $R$ as follows
\begin{align}
R(x):=1-e^{-S(x)}.
\label{eq:siam:uniquenessOfContinuousValueFunctions:Kruskov}
\end{align}
From the transform theorem for viscosity solutions \cite[Prop 2.5, Ch.2]{Bardi} it follows  that
$S$ is a viscosity solution of
Eq~\eqref{eq:siam:uniquenessOfContinuousValueFunctionsmintime:HJB}
iff the function $R$ is a viscosity solution of the equation
\begin{align}
\mathop {\sup }\limits_{v \in \controlValueSet} \left\{ { -
DR(x)\cdot f(x,v) - 1+R(x)} \right\}
=0,\\
\Leftrightarrow\,\,R(x)+\mathop {\sup }\limits_{v \in
\controlValueSet} \left\{ { - DR(x)\cdot f(x,v) - 1} \right\}
=0.\label{eq:siam:uniquenessOfContinuousValueFunctions:HJBforR}
\end{align}
As $R$ is continuous (due to the continuity assumption on $S$), from  Lemma \ref{lem:siam:SisaContinViscSoln} we have that $R$ is a viscosity solution 
to the HJB equation \eqref{eq:siam:uniquenessOfContinuousValueFunctions:HJBforR}.  Moreover, this new HJB equation, denoted by $\tilde{F}(x,DR,R)=0$ (say), has the
$\tilde{F}$ expression satisfying the requirements of
Theorem~\ref{lem:siam:generalUniquenessViscResult}.
Hence $R$ is a unique viscosity solution to
Eq~\eqref{eq:siam:uniquenessOfContinuousValueFunctions:HJBforR}
with boundary condition $R(x)=0\,,\,\forall\,x\,\in\,\targetSet$.
As the Kruskov transform is a diffeomorphism,
it follows that the minimum time function is the unique viscosity solution to the HJB
Eq~\eqref{eq:siam:uniquenessOfContinuousValueFunctionsmintime:HJB}.
\end{proof} 

The corresponding result for an undiscounted cost function  with a generalized running cost $\ell(x,v)$ is  now described.

\begin{thm}[Existence and Uniqueness]
\label{siam:thm:Suniquelam0VSgeneral}
Assume that the value function $S$  (with $\lambda =0$) is continuous in $\lieG \setminus \targetSet$. In addition assume that 
\begin{enumerate}
\item $\ell: \lieG \times \controlValueSet \rightarrow \mathbb{R}$ is uniformly Lipschitz continuous and bounded.
\item $f:\lieG \times \controlValueSet \rightarrow T\lieG$ is uniformly Lipschitz continuous, bounded.
\end{enumerate}
then $S$ is the unique viscosity solution of 
\begin{align}
H(x,DS) = 0, \quad \text{ in } \lieG \setminus \targetSet, \label{eq:siam:hjb1}
\end{align}
and $R(x)=1-e^{-S(x)}$ is the unique viscosity solution of 
\begin{align}
R(x)+ \mathop {\sup }\limits_{v \in \controlValueSet} \left\{ { -
D R(x)\cdot f(x,v) - \ell(x,v)} \right\},\quad \text{in}\,\, \lieG \setminus \targetSet. \label{eq:siam:hjbreq}
\end{align}

\end{thm}
\begin{proof}
By  a generalization of \cite[Ch.4, Prop. 3.12]{Bardi} to the Lie group setting,  $R$ is a viscosity solution 
of Eq.~\eqref{eq:siam:hjbreq} and, using the transformation  theorem for viscosity solutions \cite[Prop 2.5, Ch.2]{Bardi}, $S$ is a viscosity solution of Eq.~\eqref{eq:siam:hjb1}.
The uniqueness follows as the forms of these Hamiltonians satisfy the conditions in  Lemma~\ref{lem:siam:generalUniquenessViscResult}.

\end{proof}

\section{ Approximating the Viscosity solution  on Lie groups via Triangulations}
\label{sec:siam:discretization}

In this section we describe the main focus of this work on  numerical methods based on discretizing the HJB equation. This  requires a discretization of the Lie group on which the system evolves. The intuitive idea is to obtain a grid on the Lie group. This is a generalization of the idea of triangulating a 2-dimensional surface.  Such triangulations are used in numerical approximation procedures to obtain a solution to a discretization of the problem \cite{kushner1992nms}. In the latter part of this section we apply  the notion of viscosity solutions to the HJB equation, to prove the validity of such numerical approximations.

\subsection{Setup for Discretization}
In order to describe the discretization we recall some definitions from  \cite[Section 7.1, Ch.II ]{munkres1973edt}:
\begin{definition}[Simplex]
If $v_0$, $v_1\,\ldots v_m$ are independent points of $\mathbb{R}^n$, the simplex $\sigma$ which they span is the set of points $x$ such that $x=\Sigma b_i \, v_i$. Note that $b_i\geq0$ and $\Sigma b_i=1$. The numbers $b_i$ are termed the \textit{barycentric coordinates} of $x$ and, due to the properties listed here, may be used as transition probabilities in the  algorithms described below.
\end{definition}
\begin{definition}[Face of a Simplex]
A face of the simplex $\sigma$ is the simplex spanned by a subset of the vertices of $\sigma$.
\end{definition}
\begin{definition}[Simplical Complex]
A simplical complex $\mathrm{K}$  is a collection of simplices in    $\mathbb{R}^n$ such that 
\begin{enumerate}
\item Every face of the simplex of $\mathrm{K}$ is in $\mathrm{K}$.
\item The intersection of two simplices of $\mathrm{K}$ is a face of each of them. 
\item Each point of $|\mathrm{K}|$ has a neighborhood intersecting only finitely many simplices of $\mathrm{K}$ where $|\mathrm{K}|$ denotes the union of simplices of $\mathrm{K}$.
\end{enumerate}
\end{definition}
\begin{definition}[Star]
If $x$ is a point of $|\mathrm{K}|$, the \it{star} of $x$ in $\mathrm{K}$ is the union of the interiors of all the simplices $\sigma$ such that $x$ lies in $\sigma$. It is denoted by $St(x,\mathrm{K})$.
\end{definition}

Let $\omega$ be a $C^r$ map. Given a point $b$ in $\sigma$ we define the map $d \omega_{b}:\sigma \rightarrow \mathbb{R}^n$ as
\begin{align}
d\omega_{b}(x):=D\omega(b)\cdot (x-b).
\end{align}
We now recall the definition of a triangulation  \cite[Section 8.3, Ch.II ]{munkres1973edt} , which is the main concept required for the discretization of the group.
\begin{definition}[Triangulation]
A  $C^{r}$ map $\omega: \mathrm{K}\rightarrow \lieG$ is said to be an immersion if  
$d \omega _x : \overline{St(x,\mathrm{K})}\rightarrow \mathbb{R}^n$
is injective for each $x$. Such an immersion which is a homeomorphism onto, is called  a $C^r$ triangulation of $\lieG$.
\end{definition}

\begin{figure}[t!]
\centering
\includegraphics[scale=.25]{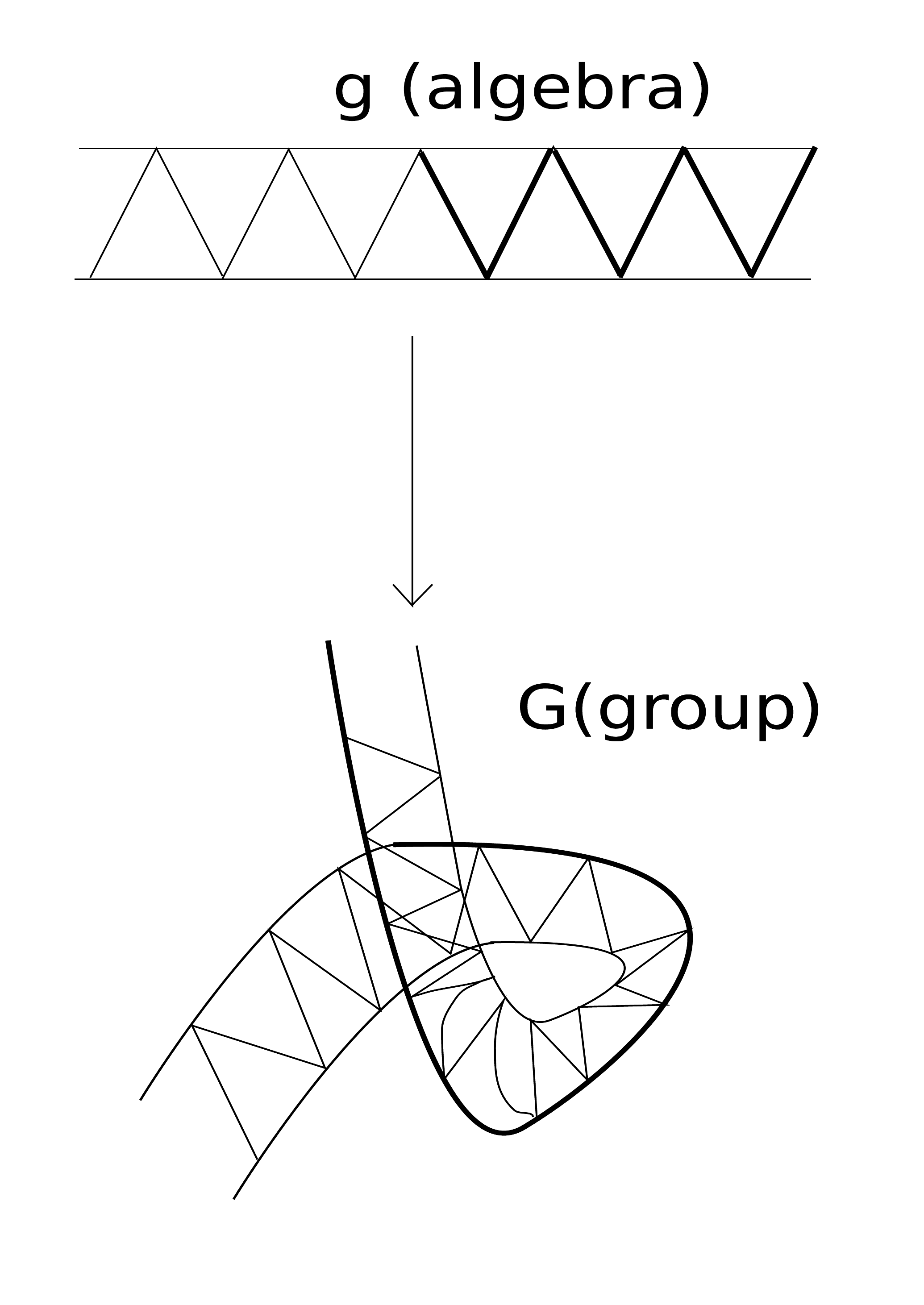}
\caption{Mapping (with overlaps) from the algebra to the group.}\label{fig:siam:overlapOfcomplexes}
\end{figure}

From \cite[Theorem 10.6]{munkres1973edt} it follows that every differential  manifold has a triangulation. Additionally, since our problem framework is on a compact connected Lie group there is a set of natural imbeddings from complexes in the Lie algebra $\lieg$ to the group via the exponential mapping. 
Note that complexes in the algebra may be easily obtained in this case, for instance via triangulations on the algebra (since the algebra in our case is isomorphic to some Euclidean space). Note that there may be overlaps between the images (on the Lie group) of these complexes. For instance we may have $C^r$ imbeddings of two complexes into $\lieG$ whose images overlap. It can be shown that by altering these imbeddings slightly their images can be made to fit together \lq nicely\rq\, i.e intersect in a subcomplex after suitable subdivisions. The intuition behind this is conveyed in Fig \ref{fig:siam:overlapOfcomplexes}. This is an especially important concept since mappings such as the exponential map are not injective and carry multiple points in the algebra to the same point on the group. 

The following notations are used in this section:
\begin{itemize}
  \item $\mathbf{L}^{\circ}:=\lieG \setminus \targetSet$.
  \item $\mathbf{L}:=$ closure of $\mathbf{L}^{\circ}$.
  \item $S_h$: simplex on $\lieG$ with maximum distance between points being $h$.
  \item $\mathbf{L}^{\circ}_h:=S_h\bigcap \mathbf{L}^o$.
  \item $\partial  \mathbf{L}^{\circ}_{h}$: subset of $\mathbf{L}^{\circ}_h$ s.t
  \begin{align}
\mathop {\sup }\limits_{x\in \partial  \mathbf{L}^{\circ}_{h}} \mathop {\inf
}\limits_{y\in \partial \targetSet} \left\| {x - y} \right\| &\rightarrow
0\qquad\mathrm{as}\,\, h\rightarrow 0. 
  \end{align}
Thus  $\partial  \mathbf{L}^{\circ}_{h}$ is the approximation to the boundary of $\mathbf{L}^{\circ}$.
\end{itemize}
We recall that the system is denoted by
\[
\dot{x}=f(x,v)\,,\qquad x\,\in\,\lieG
\]
where $v$ is a control signal chosen from a compact topological space
of controls $\controlSet$.

\begin{figure}[t!]
\centering
\includegraphics[scale=.4]{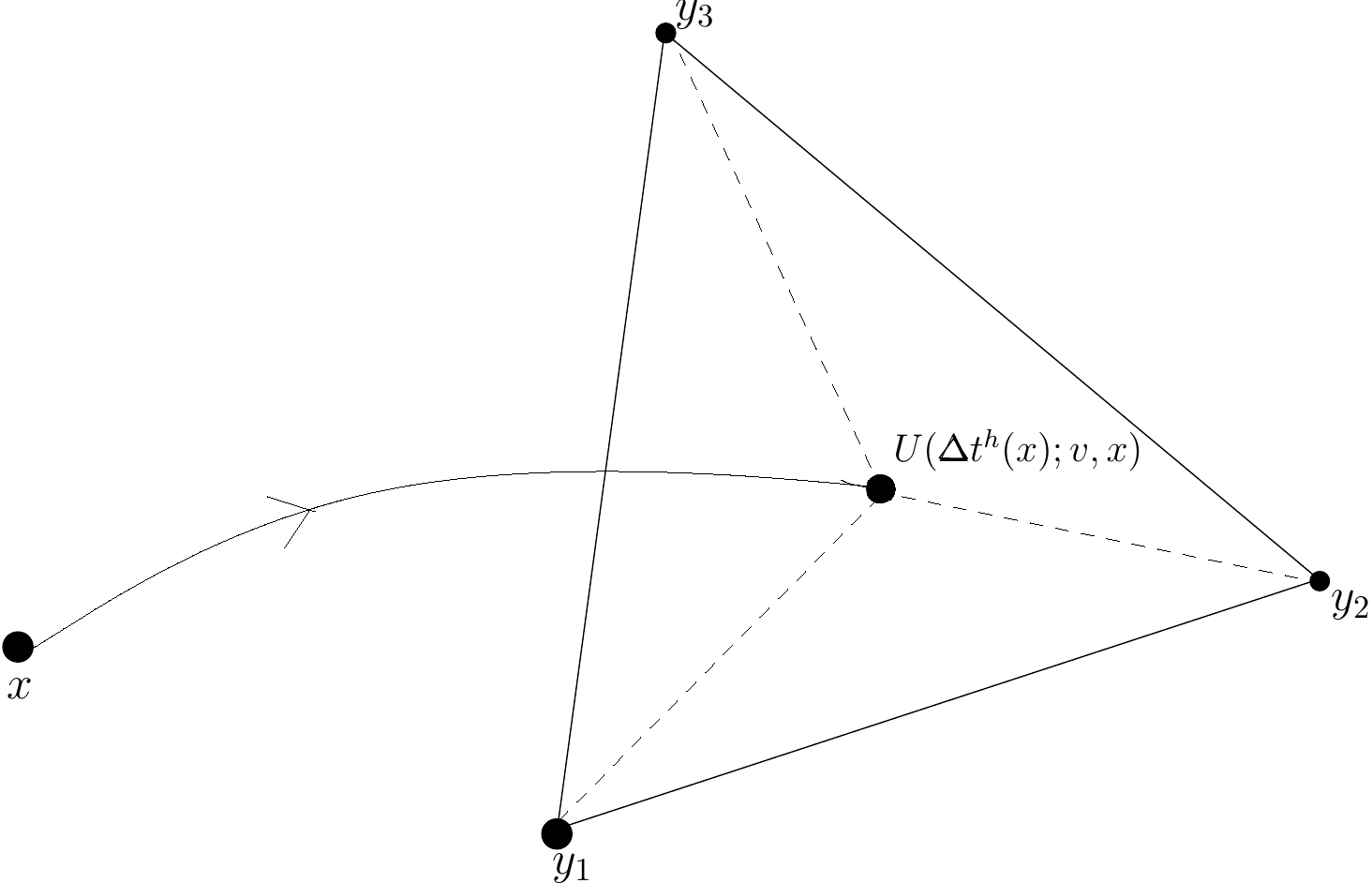}
\caption{Discretization using Barycentric coordinates.}\label{fig:siam:discretizationpic}
\end{figure}

Now consider the continuous viscosity solution $V$ to the HJB equation
 on the set $\mathbf{L}$. The discretized version of the
solution is denoted by $V^h$. In the case of the exponentially discounted (with discounting factor $\lambda$), infinite
time horizon control problem  \cite{Bardi,kushner1992nms}  the HJB equation is
\begin{align}
\lambda V + \mathop {\sup }\limits_{v \in \controlValueSet} \{ -
DV(x)\cdot f(x,v) - \ell(x,v)\}&=0 , \label{equation:siam:HJBeqndisc} \\
\mathrm{with\,boundary\,condition}\nonumber\\
V(x)=0,\qquad \,x\in\,\partial \targetSet. \nonumber
\end{align}
In order to discretize the HJB equation we define the following terms 
\begin{align}
B(x) &:= \sup_{v \in \controlValueSet} \|f(x,v)\|, \\
\mathrm{and} \quad \deltatx &:= \frac{h}{B(x)}.
\end{align}
and use the following discretization for $DV\cdot f$
\begin{align}
& DV\cdot f(x,v) =  \nonumber \\ &\frac{1}{h} \, \left \{\sum_{y \in \Sigma^h}{ p(x,y|v) V(y)} - V(x) \right \} \cdot f(x,v).
\end{align}
Here $\Sigma^h(x)$ is the set of vertices of the simplex which contains the point arrived at by flowing for time $\deltatx$  along $f(x,v)$ using a constant control signal $v$ starting from the point $x$. This is depicted in Fig.\ref{fig:siam:discretizationpic}.  For each value of $v$ the terms $p(x,y|v)$ must satisfy 
 $$
\sum_{y\in \Sigma^h(x)} p(x,y|v)=1,
 $$
and can therefore be interpreted as transition probabilities. Hence they can be obtained naturally from the Barycentric co-ordinates  on the complex as mentioned previously.  Using this discretization in the HJB equation \eqref{equation:siam:HJBeqndisc} and rearranging we obtain
\begin{align}
 &\lambda V^{h}(x) = \nonumber \\ 
 &\inf_{v\in \controlValueSet} \Biggl{\{} \frac{1}{h} \,\Big{[} \sum_{y \in \Sigma^h}{ p(x,y|v) V(y)} - V(x) \Big {]} \cdot f(x,v)   \nonumber \\  & \qquad   + \ell(x,v) \Biggr{\}} .
\end{align}

Dividing throughout by $B(x)$ and rearranging, we get
\begin{align}
& \frac{\lambda V(x) h}{B(x)} = \nonumber \\& \inf_{v\in \controlValueSet}  \frac{1}{B(x)} \Biggl{\{} \,\Big{[} \sum_{y \in \Sigma^h}{ p(x,y|v) V(y)} - V(x) \Big {]} \cdot f(x,v)  \nonumber \\ &  \qquad \qquad + \ell(x,v) h \Biggr{\}}, 
\end{align}
which implies
\begin{align}
&V(x) \left (1+\frac{\lambda  h}{B(x)}\right ) \nonumber \\  &= \inf_{v\in \controlValueSet}  \left \{ \, \frac{\sum_{y \in \Sigma^h}{ \{p(x,y|v) V(y) f(x,v)\}}}{B(x)}  \right .
\nonumber \\ & \left. 
+ V(x) \Big{(}1-  \frac{f(x,v)}{B(x)} \Big{)}+ \frac{\ell(x,v) h}{B(x)} \right \}.
\nonumber
\end{align}
Therefore,
\begin{align}
& V(x) \left (1+ \lambda \deltatx \right ) 
  \\ &= \inf_{v\in \controlValueSet} \Bigl{\{} \, \frac{\sum_{y \in \Sigma^h}{ \{p(x,y|v) V(y) f(x,v)\}}}{B(x)} \nonumber \\ 
  & + V(x) \Big{(}1-  \frac{f(x,v)}{B(x)}\Big{)}+ \ell(x,v) \deltatx \Bigr{\}}.
  \nonumber
\end{align}
We define a new set of transition probabilities 
\begin{align}
\tilde{p}(x,y|v) &= p(x,y|v)\,\frac{f(x,v)}{B(x)}, \quad \forall y \neq x\\
\tilde{p}(x,x|v)  &= 1-\frac{f(x,v)}{B(x)}.
\end{align}
Hence the discretization of the HJB equation can be shown to satisfy
\begin{align}
 & V^{h}(x)= \nonumber \\
 & \mathop {\inf }\limits_{v\in \controlValueSet}
\left\{ \bigg{(}{\sum\limits_{y\in \tilde{\Sigma}^{h}(x)} {\tilde{p}(x,y|v)}
V^{h}(y) }\right . \nonumber \\ &  { + \ell(x,v) \deltatx} \bigg{)} \times
\frac{1}{1+\lambda \deltatx}
\Biggr{\}},\label{eq:writeup:defineVh}
\end{align}
with boundary condition
$$
V^{h}(x)=0,\qquad \,x\in\,\partial \mathbf{L}^{\circ}_{h},
$$
where
\begin{align}
\tilde{\Sigma}^{h}(x):=  \Sigma^{h}(x)\bigcup {\{x\}}.
\end{align}

We rewrite  Eq~\eqref{eq:writeup:defineVh} in the more general form
\begin{align}
V^h(x)=F^{h}(V^h)(x). \label{eq:siamj:contractionMapDef}
\end{align}
Hence it can be seen that $V^h$ is
the fixed point of an operator $F^h$. We assume that there exists a parameter $\gamma_h \in (0,1]$ which is continuous with respect to $h$ with the property  that $\mathop {\lim
}\limits_{h \to 0} \gamma_h  = 1$.  It can be checked  that  $F^h(\cdot)$ has
 the following  properties:
\begin{enumerate}
\item For all constant valued
functions $c$
$$F^h(V^h+c)(x)=F^h(V^h)(x)+\gamma_{h}^{-1}c\,,\quad x\,\in\,\mathbf{L}^{\circ}_{h}.$$ 
\item  For all $\omega\in C^{1}(\mathbf{L})$
\begin{align}
\begin{array}{l}
 {\mathop {\lim }\limits_{ y \to x \hfill \atop
  h \downarrow 0 \hfill}\frac{{F^h(\omega)(y) - \omega(y)}}{h} \ge 0} \, \Leftrightarrow\, H(x,D \omega, \omega) \le 0 \label{eq:writeupFlatspace:limitOfFtoHJB}\\
 {\mathop {\lim }\limits_{ y \to x \hfill \atop
  h \downarrow 0 \hfill} \frac{{F^h(\omega)(y) - \omega(y)}}{h} \le 0}\,  \Leftrightarrow \, H(x,D \omega, \omega) \ge 0 \\
 \end{array}
\end{align}
where $H(\cdot)=0$ is the HJB equation  for the problem.
\end{enumerate}
Note that these assumptions hold for the particular choice of discretization via transition probabilities that we have outlined. Under these assumptions we proceed to look at  convergence  results for certain approximations to the value function.

\subsection{Convergence of the Approximation}
In this section we prove the validity
of the approximations to the HJB.  Our aim is to prove the convergence of  the two terms defined below to the viscosity solution.
\begin{align}
V^*(x)&:= \mathop {\lim }\limits_{\scriptstyle y \to x \hfill \atop
  \scriptstyle h \downarrow 0 \hfill} \sup V^{h}(y),\\
  V_{*}(x)&:= \mathop {\lim }\limits_{\scriptstyle y \to x \hfill \atop
  \scriptstyle h \downarrow 0 \hfill} \inf V^{h}(y).
\end{align}

The following is a generalization of  \cite[Ch.IX, Lemma
4.1,Theorem 4.1]{fleming2006cmp} 
\begin{lemma}
$V^{*}$ is a viscosity sub-solution to the HJB $H(x,DV,V)=0$.
\end{lemma}
\begin{proof}
Let $\omega\,\in\,C^{1}(\mathbf{L})$ s.t $V^{*}-\omega$ has a local maximum
at $\bar{x}\,\in\,\mathbf{L}$.  Note that without loss of generality we can assume that $\bar{x}$ is a strict maximum on $\mathbf{L}$ by redefining $\omega$ suitably \cite{kushner1992nms}.
There exists some subsequence $h$  converging to zero s.t
$V^h-\omega$ has a maximum at $y_h$  on $\mathbf{L}_{h}$ (such that
$y_h\rightarrow \bar{x}$ as $h$ goes to zero). From the local maximum
property it follows that
\begin{align}
V^h(y_h)-\omega(y_h)&\geq V^h(y)-\omega(y)\,,\quad
y\,\in\,\mathbf{L}^{\circ}_{h}.
\end{align}
Hence taking a $\gamma_h\,\in\,(0,1]$ we have
\begin{align}
\gamma_{h}^{-1}\,( V^h(y_h)-\omega(y_h) )& \geq
 V^h(y)-\omega(y)\,,\quad y\,\in\,\mathbf{L}^{o}_{h}\\
\Rightarrow\;\;\omega(y)-{\gamma_{h}}^{-1}\omega(y_h)&\geq
V^h(y)-{\gamma_{h}}^{-1}V^h(y_h).
\end{align}
Applying the operator $F^h$ and using the properties that it
satisfies we obtain
\begin{align}
F^{h}(\omega)(y_h)-\omega(y_h)\geq F^{h}(V^h)(y_h)-V^{h}(y_h).
\end{align}
At $y=y_h$ the right hand side of the above equation is zero (from Eq\eqref{eq:siamj:contractionMapDef}).
Dividing by $h$ and taking the limit as $h$ tends to zero, we have
\begin{align}
{\mathop {\lim }\limits_{y_h \to \bar{x} \hfill \atop
  h \downarrow 0 \hfill} \frac{{F^{h}(\omega)(y_h) -
\omega(y_h)}}{h} \ge 0}\Leftrightarrow \nonumber\\
{\mathop {\lim }\limits_{h \to 0} \frac{{F^{h}(\omega)(\bar{x}) -
\omega(\bar{x})}}{h} \ge 0}
\end{align}
Using Eq~\eqref{eq:writeupFlatspace:limitOfFtoHJB} we obtain
\[
H(x,D\omega,\omega)\leq0.
\]
Hence the theorem is proved.
\end{proof}
Similarly we can show that $V_{*}$ is a viscosity super-solution to the HJB equation.  We then have the following result.

\begin{theorem}
Assume that $V^h$ converges to the boundary conditions in a uniform manner i.e
\begin{align}
\mathop {\lim }\limits_{\scriptstyle y \to x \hfill \atop
  \scriptstyle h \downarrow 0 \hfill} V^h(y) = V(x)\,,\qquad
  x\,\in\,\partial  \targetSet.
  \label{eq:writeup:unifReachingBoundaryConditions}
\end{align}
We then have
$$\mathop {\lim }\limits_{\scriptstyle y \to x \hfill \atop
  \scriptstyle h \downarrow 0 \hfill} V^{h}(y)=V(x),\qquad x\,\in\,\mathbf{L}.$$
\end{theorem}
\begin{proof}
$V_*$ was shown to be a viscosity  super-solution to the HJB. Hence
using Eq~\eqref{eq:writeup:unifReachingBoundaryConditions} we have
from standard comparison theorems for continuous viscosity solutions
\cite{Azagra2006} that
\begin{align}
V(x) \leq V_{*}(x)\,,\qquad
x\,\in\,\mathbf{L}.\label{eq:writeup:comparisonForVsubstar}
\end{align}
Similarly since it was shown that $V^*$ is a viscosity sub-solution
to the HJB, we apply the comparison theorem to obtain
\begin{align}
V(x) \geq V^{*}(x)\,,\qquad
x\,\in\,\mathbf{L}.\label{eq:writeup:comparisonForVsuperstar}
\end{align}
From
Eqns~\eqref{eq:writeup:comparisonForVsubstar},\eqref{eq:writeup:comparisonForVsuperstar}
we have that
$$
V^{*}\leq V_{*} \,\mathrm{on\;} \mathbf{L}.
$$
However from the definitions of $V^*$ and $V_*$ we have that
$$
V^{*}\geq V_{*} \,\mathrm{on\;} \mathbf{L}.
$$
Hence using the two inequalities above we have $V^*=V_*$ on $\mathbf{L}$ and the statement of the theorem follows.
\end{proof}

\section{Simulations}\label{sec:siam:simulation}
We now provide an example problem on $SU(2)$ to demonstrate the results of this article. The 
system dynamics is given by
\begin{align}
   \dot{U}  &=  [ v_1\, I_x+v_2\,I_z  ]U,\quad\,\|v\|=2 \label{equation:siam:simExampleSystem}  \\
   U(0)  &= U_0, \nonumber \\
   U_0\,,\,U\,\in\,SU(2).\nonumber
\end{align}
The value function for this problem is given by
\begin{align}
S(U_0)&= \mathop {\inf}\limits_{v\in \controlSet } {\int\limits_0^{t_{U_0}(v)} {\exp\{-\lambda\,s\}\, ds}}.
\end{align}

 As previously mentioned, $\exp(\lieg)$ generates $\lieG$. Hence instead of obtaining several complexes, patching them together and refining them on the areas of overlap, we take a sufficiently large area of the algebra $\mathfrak{su}(2)$ and perform a triangulation  on it. This is followed by \lq glueing\rq\, together the value function at points corresponding to the identity element of $SU(2)$. This simplified discretization will yield accurate solutions close to the target set (and on the interior of the region being considered), but loses accuracy towards the end of the region. The mapping ${\exp(x\,I_x+y\,I_y+z\,I_z)}$  from $\mathfrak{su}(2)$ to $SU(2)$  provides three natural parameters with which to visualize the value function on the Lie algebra. Note that  the map is not injective and multiple points from the algebra may be mapped to the same point on the group.

\begin{figure}[h!]
\centering
\includegraphics[scale=.4]{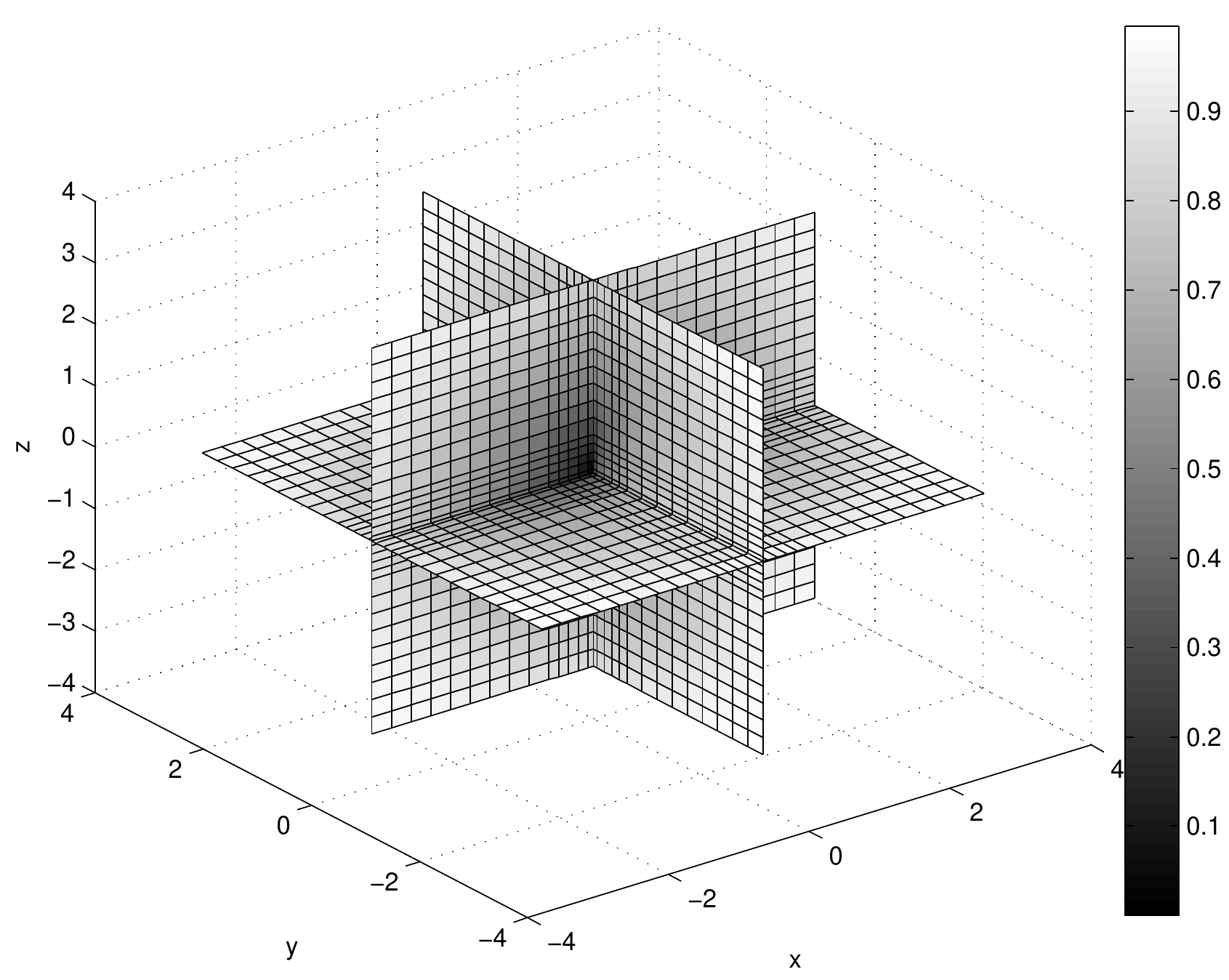}
\caption{Normalized minimum time function with a control of norm
2. }\label{fig:siam:valueiterationsu2}
\end{figure}

\begin{figure}[t!]
\centering
\includegraphics[scale=.5]{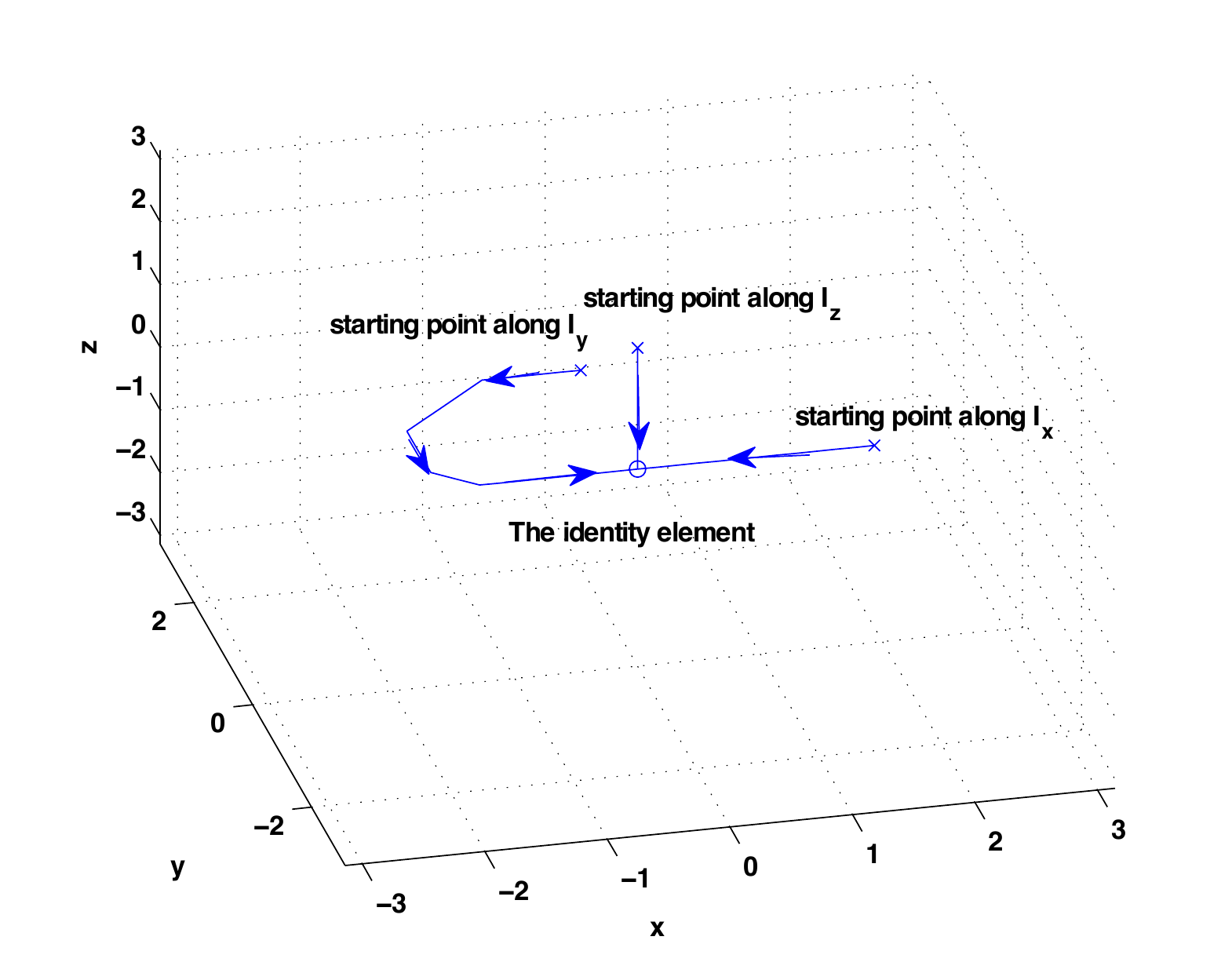}
\caption{Optimal trajectories to the identity element starting from different points on each of the axes.}\label{fig:siam:sampletrajectory}
\end{figure}

\begin{figure}[t!]
\centering
\includegraphics[scale=.44]{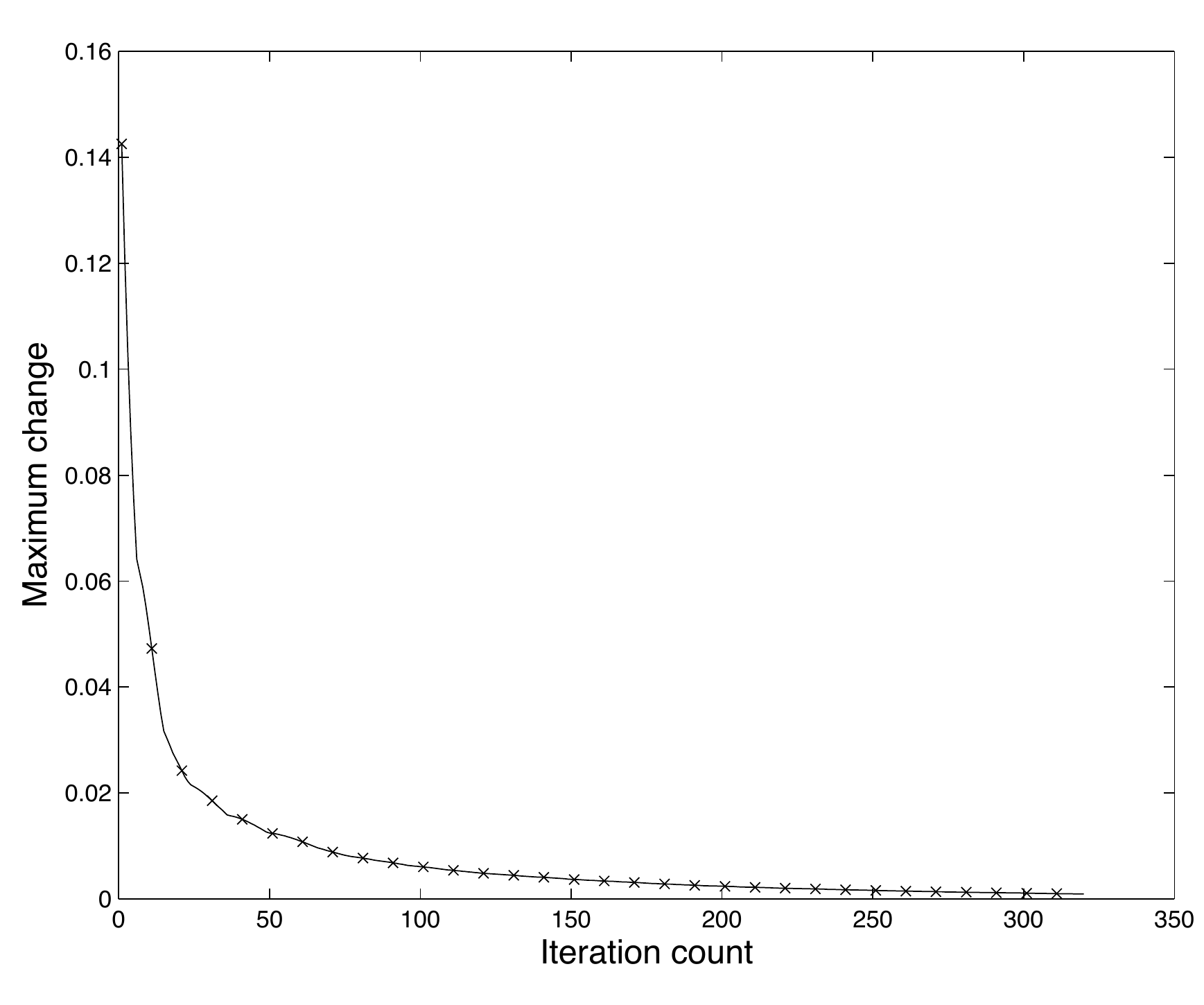}
\caption{Maximum of the absolute change in the  value function over the grid at each iteration.}\label{fig:siam:stoppingmetric}
\end{figure}

We use standard numerical tools from computational geometry to obtain a triangulation of the chosen subset of the algebra visualized as a region in $\mathbb{R}^3$ (since $\mathfrak{su}(2)$ is isomorphic to $\mathbb{R}^3$). Performing the value iteration mentioned in Section \ref{sec:siam:discretization} the resulting value function is shown in Fig \ref{fig:siam:valueiterationsu2}. Lighter shading at a point indicates a larger value of the minimum time function at that location. 
The stopping criteria used in the algorithm designed is as follows.  At each step of the iteration the absolute change in the value function at all points on the mesh is computed. The maximum value of this change across the entire mesh  is determined  and a threshold value for this stopping metric, at which the algorithm should terminate, is set.  This metric is indicated in Figure \ref{fig:siam:stoppingmetric}. Note  that there are several different possibilities choices for a stopping criteria. We defer to future work the  analysis of various possible stopping metrics and a detailed quantitative study of numerical convergence rates and error bounds.

The optimal controls are obtained using a discretized version of the verification theorem and from them we can obtain sample trajectories from various starting points as indicated in Fig \ref{fig:siam:sampletrajectory}. Note that this figure must be carefully interpreted since the mapping  from this region into $SU(2)$ is not injective.

\section{Conclusion}\label{sec:siam:conclusion}
In this paper we introduced a rigorous framework for the numerical techniques involved in using the  Dynamic Programming technique from optimal control theory for the control of quantum spin systems evolving on compact Lie groups. Numerical simulations were preformed by triangulation of the group which, due to the well studied numerical procedures available for tesselation of surfaces, enable better  numerical speed and efficiency of implementation. In addition, the solution can be made more accurate at points of the group where such accuracy is desired (e.g around the origin   in Fig.~\ref{fig:siam:valueiterationsu2} where the solution is non differentiable). The dynamic programming methods provide a framework that can, in principle, be used for systems with an arbitrary number of qubits unlike limitations on the Lie theoretic methods. In addition, alternative numerical techniques that use the calculus of variations are subjected to issues in the entrapment at local minima - a drawback absent in the current approach.

The value function iteration methods when used on any grid, suffer from the curse of dimensionality and hence become intractable for higher dimensional systems. For instance, the number of spatial dimensions in a quantum spin $1/2$ system with~$n$ qubits  grows as ${4^{n}-1}$. Possible directions of future work may involve a study of methods  such as fast marching \cite{kimmel1998cgp} or meshless techniques to improve the speed of computations. 
Inspired by the dynamic programming framework in this article and the curse of dimensionality free approaches in \cite{mceneaney2008cdf},  new methods are currently being developed for reduced dimensionality approximation techniques to quantum control. 

There exist classes of control problems such as those involving quantum systems with bounded controls and drift  for which the value function is discontinuous. Viscosity solution techniques for such  discontinuous cost functions may be used to provide the technical framework for the use of  impulsive  controls in the dynamic programming approach to quantum control.

\section{Acknowledgement}
This research was supported by the   Australian Research Council. The authors would also like to thank the anonymous reviewer for helpful comments.
\bibliographystyle{elsarticle-num}

\end{document}